\documentclass[aps,prl,twocolumn,groupedaddress,showpacs,floatfix]{revtex4}
\usepackage{graphicx}
\usepackage{dcolumn}
\newcolumntype{.}[1]{D{.}{.}{#1}}
\usepackage{float}
\usepackage{color}
\begin{document}

\title{The paramagnetic and glass transitions in sudoku}

\author{A~Williams and G.~J. Ackland}
\affiliation{ SUPA, School of Physics and Astronomy 
The University of Edinburgh, Edinburgh, EH9 3JZ, UK.}

\begin{abstract}
\noindent

We study the statistical mechanics of a model glassy system based on a
familiar and popular mathematical puzzle.  Sudoku puzzles provide a
very rare example of a class of frustrated systems with a unique
groundstate without symmetry.  Here, the puzzle is recast as
thermodynamic system where the number of violated rules defines the
energy. We use Monte Carlo simulation to show that the ``Sudoku
Hamiltonian'' exhibits two transitions as a function of temperature, a
paramagnetic and a glass transition.  Of these, the intermediate condensed phase is
the only one which visits the ground state (i.e. it solves the puzzle,
though this is not the purpose of the study).  Both transitions are
associated with an entropy change, paramagnetism measured from the
dynamics of the Monte Carlo run, showing a peak in specific heat,
while the residual glass entropy is determined by finding multiple instances of
the glass by repeated annealing.  There are relatively few such simple
models for frustrated or glassy systems which exhibit both ordering
and glass transitions, sudoku puzzles are unique for the ease with
which they can be obtained with the proof of the existence of a
unique ground state via the satisfiability of all constraints.
Simulations suggest that in the glass phase there is an increase in information entropy
with lowering temperature.  In fact, we have shown that sudoku have
the type of rugged energy landscape with multiple minima which
typifies glasses in many physical systems, and this puzzling result
is a manifestation of the paradox of the residual glass entropy.
These readily-available puzzles can now be used as solvable model
Hamiltonian systems for studying the glass transition.

\end{abstract}
\date{\today}
\pacs{}
\maketitle
%
%

Complex systems such as glasses are typified by frustrated
interactions which cannot be simultaneously satisfied.  By contrast,
systems with a well defined ground state are typified by some
translational symmetry and long ranged ordering.  The popular
mathematical puzzle sudoku lies intermediate between these two.
Standard sudokus, printed by the thousand in newspapers around the
world, have a unique solution which normally has no symmetry.  Adding
one ``wrong'' number-constraint tips the problem into the regime of the
frustrated system, which cannot be solved, while removing one of the
clues leads to a system with multiple solutions.

We obtained our sudoku puzzles from the online  Playr site\cite{playr}, and also investigate a minimal 17-clue puzzle\cite{17}
Overconstrained and underconstrained puzzles were created by adding or removing 
constraints from these puzzles.
For human solution, the puzzle should be solved by ``logic'', 
and puzzles are categorized by how difficult it is to do this.
The Playr site rates puzzles on a five-point scale. 

The sudoku problem can be cast into a Hamiltonian form.  Here a 9x9 grid 
contains ``spins'', taking integer values between 1 and 9.  The sudoku 
rules require that no ``spins'' in the same row, column or 3x3 sub-grid
should take the same value.  By assigning an energy to each violation 
of the rules, we arrive at the following Hamiltonian
\begin{equation}
H = \sum_i H(\{\sigma_i\}) =  \sum_{i}\sum_{j}\delta_{\sigma_i\sigma_j}
\label{eq:h}
\end{equation}
Where the $j$ sum runs over all sites which share a row, column, or
sub-grid with $i$.  The problem is defined by assigning fixed values
to some spins (``clues''), and treating the others as dynamic variables.  The
solution to the sudoku problem is the ground state of the Hamiltonian, which
has $H=0$. It can be seen that this is equivalent to a 9-state Potts Hamiltonian
on a 20-neighbour finite periodic lattice.\cite{potts}

Here, we define microstates of the system with each $\sigma_i$s taking
a value between 1 and 9.  We then sample the microstates 
according to the Boltzmann distribution with a fictitious temperature,
T\cite{kb}, i.e. the probability of a state with energy $H$ is
$p(H)\sim \exp{(-H/T)}$.  This is done by accepting or rejecting trial
changes of individual numbers using the Metropolis
algorithm\cite{metro}. This Markov process could be set to terminate
once $H=0$, acting as a solver, however better methods are available
for solving sudoku puzzles\cite{gunther,babu,knuth} and this is not
the purpose of this work.  We are interested in the equilibrium
macrostate and glassy dynamics so we will use the Monte Carlo
simulation to obtain ensemble averages.

\begin{figure}
\includegraphics[width=0.45\textwidth,clip]{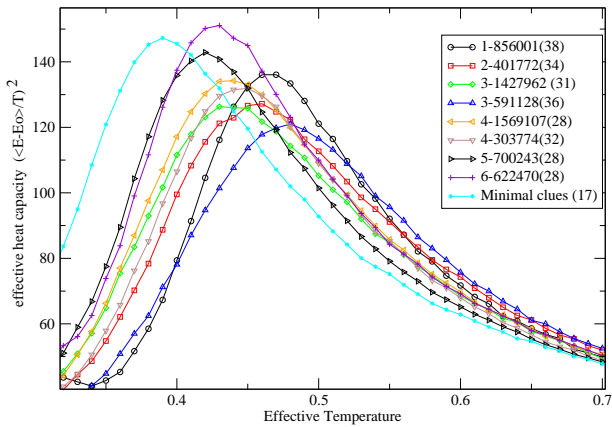}
\caption{Variation of ``heat capacity'' with temperature, showing
  that the transition temperature shifts with number of fixed 
values.  Legend shows the Playr puzzle number,  with the first digit labelling the ``difficulty'' and the number in
  brackets the number of fixed sites. \label{fig:En}}
\end{figure}

\begin{figure}
\includegraphics[width=0.45\textwidth,clip]{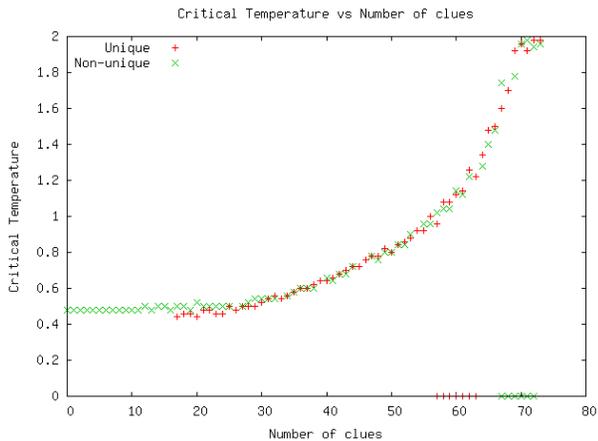}
\caption{Variation of critical temperature with number of clues for puzzles with and without unique solutions.
\label{fig:Tc}}
\end{figure}

\begin{figure}
\includegraphics[width=0.45\textwidth,clip]{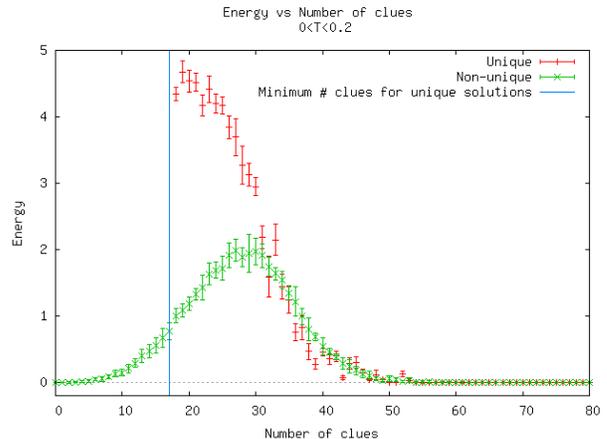}
\caption{Variation of mean energy in the $T<0.2$ range for a series of puzzles
  with different number of fixed sites, showing the appearance of 
a glassy state for intermediate numbers of clues, and an increased 
tendency towards glassiness for puzzles with a unique solution\label{fig:fixed}}
\end{figure}

The mean energy $<H>$ was evaluated,
and the specific heat 
\begin{equation} c_v = <(H-<H>)^2>/T^2\end{equation} 
for representative problems are shown in figure \ref{fig:En}.  This
shows a transition between a low temperature ``ordered'' state
close to the ground state, and a high temperature ``paramagnetic''
state with many rule violations.  The variance of the energy shows a
distinct peak at this temperature.  These are
the classic signals of a phase transition, and hereinafter we will
refer to it as such, however we note that since the system size is
fixed there is no ``thermodynamic limit''.   For a system of this 
small finite size, we expect such a rounded peak rather than a discontinuity.  
Larger sudoku puzzles with
4x4 and 5x5 grids exist, but the Potts-model sites have more
neighbours, so do these not comprise a limit.  The finite size of the
phase space means that peaks are rounded rather than sharp, and consequently we
were also unable to obtain sufficient data to calculate critical
exponents.

The transition temperature shows a strong dependence on the number of
fixed sites, but not on the existence of a unique ground state
solution (Figure \ref{fig:Tc}).  We investigated this by taking
published puzzles, and either fixing additional sites (consistent with
the solution) to increase that number, or removing fixed sites to
create systems with multiple minima, or both. 

Two phases discovered here are akin to a high entropy paramagnetic
phase, which samples the whole phase space, and an ordered condensed
phase which is close to the solution, while sampling 
other  well-defined,
energy minimum.  The ground state, while unique, has no symmetry, and
there are also other macrostates corresponding to metastable energy
minima in which the system can become stuck at low temperature.  We
refer to the phase stuck in this rough non-symmetric energy landscape
as a glass.

 We studied a number of
systems and monitored their ensemble average low T
energy after an equilibration time of 200000 MC steps(Fig \ref{fig:fixed}).  In general, this places
us in a non-ergodic regime with each simulation becoming fixed in a
local minimum.  It is therefore a kinetic rather than thermodynamic
measure, the non-zero values of $<H>$ being due to failing to find the global minimum below
the glass transition rather than thermal noise.  For puzzles with multiple solutions the maximum
complexity (as measured by mean low temperature energy) appears with
about 27 fixed sites.  With more fixed sites there is a smaller phase
space, and the ground state is quickly found.  With fewer clues,
there is an increasing number of zero energy states, and again one is
easily located.  This may be regarded as a degenerate glass.

This latter argument does not hold for puzzles with a unique solution,
where equilibration from an initial random state typically falls
into a local minimum.  Indeed we see that the low-temperature
ensemble-averaged energy for systems with unique solution increases
sharply towards 17, the minimum possible\cite{sudmin}.

\begin{figure}
\includegraphics[width=0.50\textwidth,clip]{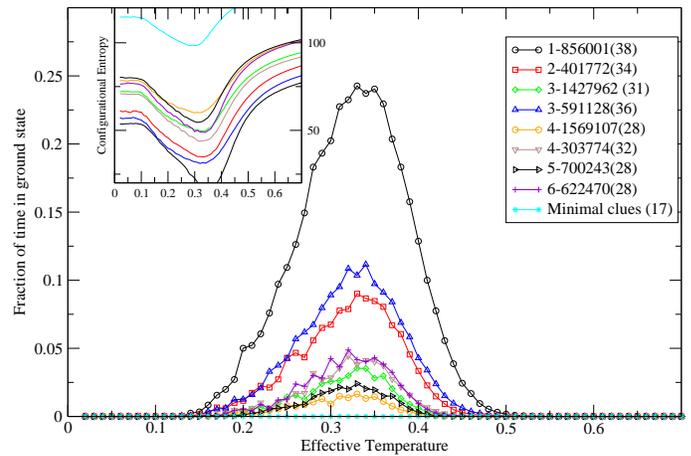}
\caption{Fraction of time spent in ground state for a series of
  puzzles with different ``difficulty'' rating, at each temperature,
  we simulated 10$^9$ attempted steps, split into 1000 groups with
  each group initially randomized, and statistics gathered over the
  final $8\times 10^5$ steps. Legend shows playr puzzle number\cite{playr}.
The solution to the 17 clue
  minimal puzzle was only found once.  Inset: $S(T)$ vs $T$.  This
  quantity is normalized to the number of sites, rather than the
  number of free sites, hence systems with a large number of free
  sites tend to have higher $S(T)$ throughout. \label{fig:glass}}
\end{figure}

The existence of a unique ground state makes it possible to be certain
when the simulation is in a metastable state close to a non-zero-energy minimum.  From this it
is also possible to probe the roughness of the energy landscape and
determine something analogous to a glass transition.  At very low
temperature the simulation falls into a local minimum and becomes
stuck.  Then on heating we enter the ``glassy'' regime where a variety of
different states are sampled, including the ground state.  This transition
is associated with a broad peak in the specific heat on the $c_v$ vs T plot, but a clearer
measure of the number of local minima can be determined by measuring
the fraction of time that the system is in the ground
state (Fig.\ref{fig:glass}).  We find that there is a correlation between both
``difficulty'' rating and number of fixed sites for the number of
metastable states: harder puzzles with fewer clues spend less time in
the global minimum, and therefore more in metastable minima.  By
contrast, the peak always falls at the same temperature, showing that
the energy barriers between the various local minima are similar in all puzzles.

The existence of a particular temperature which is optimal for finding
the ground state implies that both above and below this temperature,
the entropy of the system should be higher.  This is a more extreme
case of the residual entropy of glass and disordered
crystals\cite{glassparadox,goldstein} in that the sudoku system
appears to show a negative $dS/dT$.  Such a result, if converted from
statistical mechanical entropy to thermodynamic entropy, would imply a
violation of the second law.  However there is no physical
representation of this system, and no reversible process equating to
our simulation, so the issue is mainly one of definition of entropy.

The information (Shannon) measure is the ``configurational entropy'' of the 
system, defined as:
\begin{equation} S = -\sum_{i,j,k}^9 p(\sigma_{i,j}=k) \ln  p(\sigma_{i,j}=k) \end{equation}
where $p(\sigma_{i,j}=k)$ is the ensemble-averaged probability of
finding the number $k$ at site (i,j).  In calculating this entropy, the process for evaluating the ``ensemble average'' becomes critical.
 $S(T)$ is
relatively featureless when calculated from a single long trajectory
at each temperature, decaying monotonically to S=0 at T=0.  When
averages for $p_i$ are taken over a set of separate simulations
(i.e. repeatedly annealing to high-T) a distinct entropy minimum is
revealed at the glass transition, caused by the excess entropy from
counting multiple instances of the glass at lower temperatures. 
The variation of $S(T)$ with temperature  calculated in this way
is shown in Fig\ref{fig:glass} (inset).  It
can be seen that this minimum corresponds to the peak in finding 
solutions, and is strongly correlated with the ``difficulty'' rating 
of the puzzle: harder puzzles have more local minima.

For an abstract puzzle like sudoku there is no ``physically correct'' way of 
evaluating the ensemble average.  All we can say is that the 
ergodic hypothesis is clearly violated, and with it the link between 
statistical mechanics and thermodynamics.

To summarise, we have shown that the sudoku puzzle can be used to
define a model Hamiltonian system with interesting properties. It exhibits three phases, paramagnetic, condensed and glass  The
paramagnetic transition has a critical temperature which depends only
on the fraction of fixed sites in the system.  At low temperatures
there is a notable difference between true sudoku puzzles with unique
solutions, and the equivalent constrained Potts models with multiple
solutions: the sudoku puzzles are much stronger glass-formers.
Finally, the sudoku Hamiltonian sheds light on the residual
glass-entropy, by providing a model in which ergodicity breaking is
manifest, leading to different results for the low-temperature 
entropy depending on the method of simulation.

\begin{thebibliography}{40}
\bibitem{17} The pattern of clues in this puzzle is \newline {\small .......1......2..3...4...........5..4.16.......71.......5....2......8..4..3.91....}
\bibitem{kb} For convenience, we define units such that $k_B=1$.
\end{thebibliography}
\end{document}